**Structural and critical current properties in polycrystalline $SmFeAsO_{1-x}F_x$**

Lei Wang, Zhaoshun Gao, Yanpeng Qi, Xianping Zhang, Dongliang Wang, Yanwei Ma*

Key Laboratory of Applied Superconductivity, Institute of Electrical Engineering,

Chinese Academy of Sciences, P. O. Box 2703, Beijing 100190, China

## Abstract

A series of polycrystalline $SmFeAsO_{1-x}F_x$ bulks (x=0.15, 0.2, 0.3 and 0.4) were prepared by a conventional solid state reaction. Resistivity, susceptibility, magnetic hysteresis, critical current density and microstructure of these samples have been investigated. The resistivity result shows that critical transition temperature Tc increases steadily with increasing fluorine content, with the highest onset Tc=53 K at x=0.4. On the other hand, a study on the effect of lattice constants on critical transition temperature and resistivity at 60 K ($\rho_{60K}$) presents a possible correlation; that is, Tc rises with the shrinkage of a-axis while $\rho_{60K}$ increases with the enlargement of c-axis. A global critical current density of $1.1\times10^4$ A/cm$^2$ at 5 K in self field was achieved in the purest sample. An inter-grain Jc of about $3.6\times10^3$ A/cm$^2$ at 5 K in self field was estimated by comparing the difference in magnetization between powder and bulk samples, in contrast to the intra-grain Jc of $10^6$ A/cm$^2$. A weak field dependence was observed in the estimated inter-grain Jc, and the effect of composition gradients on the inter-grain Jc was also discussed.

* Author to whom correspondence should be addressed; E-mail: ywma@mail.iee.ac.cn

## Introduction

Since the discovery of superconductivity in the $LaFeAsO_{1-x}F_x$ compound with Tc=26 K [1], extensive efforts have been devoted to the exploration of new iron-based superconductors with higher transition temperatures. These studies have led to the discoveries of Ce, Pr, Nd, Sm, Gd iron-based superconductors [2-6]. It has been established that superconductivity can be induced in this system by the substitution of fluorine for oxygen. Oxygen vacancies under high pressure can also introduce electrons as fluorine, and result in superconductivity [7]. The hole-doping superconductor $La_{1-x}Sr_xFeAsO$ with Tc=25 K was also found in this system [8]. At present, $SmFeAsO_{1-x}F_x$ still retain the record of highest critical temperature of 55 K in iron-based oxypnictides [9].

On the other hand, many workers have focused on magnetic and transport properties of the series [10-17]. Detailed phase diagram and transport properties of $SmFeAsO_{1-x}F_x$ were studied, and it was found that superconductivity emerges at x=0.10 and optimal doping takes place at x=0.2 with the highest Tc=54 K [10]. Very high upper critical fields over 100 T have been reported for $SmFeAsO_{0.85}F_{0.15}$ and $NdFeAsO_{0.82}F_{0.18}$ [11, 12]. $LaFeAsO_{0.9}F_{0.1}$ and $SmFeAsO_{1-x}F_x$ wires, with Tc=25 K and 52 K respectively, have been prepared by our group, but low critical currents have been detected [14]. It has been proved that global critical current over the whole bulk sample does exist, and the current density is two orders of magnitude lower than that in grains. However, whether this difference is intrinsic or extrinsic is still under discussion [15]. Actually, no systematic work on magnetic and transport properties of a set of samples with various compositions has been reported so far.

In this study, we report the common features of a set of $SmFeAsO_{1-x}F_x$ (x=0.15, 0.2, 0.3 and 0.4) and present a possible relationship between Tc and lattice constants. We then estimate the inter-grain Jc by comparing the difference in magnetization between powder sample and bulk samples. We also discuss the effect of composition gradients on inter-grain current density.

## Experimental details

The polycrystalline $SmFeAsO_{1-x}F_x$ bulks were synthesized by the conventional solid state reaction. Stoichiometric amount of SmAs, $SmF_3$, FeAs and $Fe_2O_3$ powders and Sm filings were thoroughly ground in Ar atmosphere and pressed into pellets. The pellets were wrapped in Ta foil, sealed in an evacuated quartz tube, and then, sintered at 1150 ℃ for 45 hours. The end samples are dark gray and in good shape. The binary compounds SmAs and FeAs were pre-sintered in evacuated quartz tubes using Fe powder, Sm filings and As pieces as starting materials at 500 ℃ for 10 hours and 900 ℃ for 24 hours. Extra 10% As was added to compensate the evaporation of As at high temperatures.

Resistivity measurements were carried out by the standard four-probe method using a Quantum Design PPMS. Phase identification was characterized by powder X-ray diffraction (XRD) analysis with Cu-Kα radiation from 20 to 80°. Lattice constants derived from X-ray diffraction patterns were calculated using a X'Pert Plus program. Microstructure observations were performed using scanning electron microscope (SEM/EDX). Magnetization of the samples was performed using a MAGLAB and a Quantum Design SQUID. Critical current densities were calculated using Bean model.

**Results and discussion**

The powder X-ray diffraction patterns of $SmFeAsO_{1-x}F_x$ samples shown in Figure 1(a) indicate a SmFeAsO major phase and a tiny amount of SmAs and SmOF impurity phases. It is observed that, impurity phases, especially SmAs phases, increase with increasing nominal composition of fluorine. The $SmO_{0.8}F_{0.2}FeAs$ is the purest sample in the series. Figure 1(b) shows the lattice constants as a function of starting composition. A-axis lattice constant drops steadily with increasing fluorine content, however, c-axis lattice constant decreases at first, reaches its minimum value around x=0.3, and then increases with further fluorine doping.

Figure 2(a) shows the temperature dependence of resistivity for different samples. Onset Tc increases with increasing fluorine content, and reaches a maximum value of 53 K at x=0.4. For the $SmFeAsO_{0.85}F_{0.15}$, the trace of anomalous peak associated with the spin density wave (SDW) transition near 150 K can be seen. While for the other three samples, a linear temperature dependence of resistivity is observed from 60 to 300K, and the anomalous peak is totally suppressed. The RRR=ρ(300K)/ρ(60K) for samples with x=0.15, 0.2, 0.3 and 0.4 are 2.8, 4.3, 3.9 and 4.3, respectively. Figure 2(b) presents the variation of $Tc_{onset}$ and $\rho_{60K}$ with x. We can see that $Tc_{onset}$ increases steadily with x, from 39 to 53 K. While resistivity at 60 K ($\rho_{60K}$) shows a valley shape dependence on fluorine content, with a minimum value at x=0.3. It should be noted that the variation of $\rho_{60K}$ is similar to that of c-axis.

Figure 3 shows the temperature dependence of DC magnetization for the sample with the highest Tc (x=0.4). The magnetization in zero-field-cooled and field-cooled state is measured under a magnetic field of 20 Oe. Clearly, superconducting contribution is isolated and presented in the figure. Magnetic onset Tc is situated at 52 K, which is consistent with the resistivity result. The existence of superconducting phase was confirmed by the Meissner effect on cooling in a magnetic field.

Figure 4(a) and (b) show the scanning electron micrographs of the cross-section of polycrystalline $SmFeAsO_{0.8}F_{0.2}$. It can be observed that the sample is dense with some voids and the grains are well connected. In comparison to bulks synthesized by high pressure [15], our samples are more pure but less dense. High magnification images (Fig.4b) further revealed that, most grains are plate-like and in an average size of ~10μm, which were identified as SmFeAsO by EDX analysis (Fig.4d).

Magnetic hysteresis loops of $SmFeAsO_{1-x}F_x$ samples at 5 K and global critical current densities (Jc) are shown in Fig.5a and b, respectively. These hysteresis loops are all very large. No obvious paramagnetic or ferromagnetic background was observed for the x=0.2 sample, whereas for the x=0.3 and x=0.4 sample, paramagnetic and ferromagnetic backgrounds exist and seem to rise with the increase of impurity phases. Note that the x=0.2 sample provides the largest hysteresis loop. The global Jc shown in Fig.5b are extracted from the hysteresis loop widths using Bean model. For a long sample with rectangular cross-section $a \times b$ (with $a<b$), it is convenient to use the expression $Jc=20\Delta m/va(1-a/3b)$, where if $\Delta m$ is measured in emu, $v$ in $cm^3$, $a$ and $b$ in cm, $Jc$ is in $A/cm^2$. For the x=0.2 sample, the Jc is in an order of $10^4$ $A/cm^2$ at 5 K in self field, which is quite comparable to that of the samples prepared by high pressure synthesis [15]. As the external field increases, Jc decreases and then keeps constant $2\times10^3$ A/cm$^2$ over 4 T. Assuming that all current loops are restricted within grains in an average size <$R$> of 10μm, the intra-grain Jc values were calculated on the basis of $J_c = 30\Delta M/\langle R\rangle$ to be in the order of $10^6$ A/cm$^2$ at 5 K in self field, about two orders of magnitude larger than the globe Jc above.

It is known that, both the superconducting current within grains (intra-grain currents) and across the grain boundaries (inter-grain currents) contribute to the magnetization of the bulk sample, global Jc deduced from Bean model give an overestimate for the inter-grain current, which flow throughout the whole sample and can be directly measured by the resistive method. Assuming that the particles in powder sample are electrically isolated from each other, thus, the magnetization of the powder sample come from the superconducting current within each particle. Therefore, the difference in magnetization between the powder sample and the bulk sample is attributed to the superconducting currents that flow across the grain boundaries [19,11]. If the part contributed by inter-grain current loops is defined as,

$$\Delta M^{inter\text{-}grain} = \Delta M^{bulk} - \Delta M^{powder}$$

Then inter-grain current density can be estimated using Bean model.

To investigate the inter-grain Jc that flows across the grain boundaries, two rectangular specimens of equal dimension were cut from the $SmFeAsO_{0.8}F_{0.2}$ bulk sample. One of bulks was ground into powders and the magnetization loops was measured. Magnetization measurements for both the powder and bulk specimens were carried out. Figure 6(a) shows the normalized hysteresis loops of $SmFeAsO_{0.8}F_{0.2}$ bulk and powder at 5 K. Obviously, the hystersis loop width of the bulk is larger than that of the powder. This difference decreases as external fields increase, and becomes less over 4 T. The inter-grain current density for the $SmFeAsO_{0.8}F_{0.2}$ bulk derived from the expression above is about $3.6\times10^3$ A/cm$^2$ at 5 K and $8\times10^2$ A/cm$^2$ at 20 K in self field, both of which show a weak dependence of fields.

In order to get more information about the microstructure of samples, SEM and EDX mapping experiments were performed, as shown in Fig.7. All elements were detected in an appropriate ratio, indicating that the major phase is SmFeAsO. The distribution of iron and arsenic is homogeneous in the whole area. However, the deficiency of samarium, fluorine and oxygen can be seen in some areas. The inhomogeneity, or composition gradient, is thought to have a large effect on the current density of $SmFeAsO_{1-x}F_x$, which will be discussed in details later.

The substitution of fluorine for oxygen, which introduces electrons and causes distortions of crystal lattice in iron based oxypnictides, plays an important role in superconductivity. It was reported that optimal doping in $SmFeAsO_{1-x}F_x$ takes place at x=0.2 with the highest Tc=54 K [10]. However, our result shows that, critical transition temperature increases steadily with increasing fluorine content, and reaches its maximum of 53 K at x=0.4, which is different from the previous study. Recently, Takahashi et al. have reported that increasing the pressure causes a step increase in onset Tc in F-doped LaFeAsO [18]. They believed that the anisotropic shrinkage of the lattice constants introduced by external pressure is the main reason for the dramatic rise in Tc, and further, the application of external pressure increases Tc by enhancing charge transfer between the insulating and conducting layers. However, nothing about the correlation of Tc with a-axis or c-axis lattice constant has been reported.

Fluorine doping can also results in shrinkage of both a and c axes, which can be called 'chemical pressure'. Our study on this 'chemical pressure' provides more information about the correlation. From the inset of Fig.1 and the inset of Fig.2, it appears that when a-axis and c-axis shrink with the fluorine content increasing from 0.15 to 0.3, onset Tc increases from 39 to 49 K. The decrease in $\rho_{60K}$ suggests that carrier density is enlarged. Thus, the rise in onset Tc can be explain by the effect of shrinkage of lattice constants, or the enlargement of carrier density, or the interrelation of both.

With further increasing nominal fluorine content from 0.3 to 0.4, a-axis shrinks a length of about $1\times10^{-4}$ nm and c-axis increases a length of about $5\times10^{-4}$ nm. Onset Tc rises from 49 to 53 K, while carrier density decreases. It indicates that the effects of c-axis distortion and carrier density on Tc are not remarkable, while the shrinkage of a-axis seems the main reason for the increase of Tc and the change of c-axis has a large effect on carrier density. As the nature of the pairing mechanism in these systems is not understood, the empirical correlation of Tc with lattice constants can help in clarifying the situation.

One may notice that the global critical current density (Fig.5b) is almost independent of magnetic fields, while the inter-grain Jc (Fig.6b) shows a weak dependence on magnetic fields. A possible explanation about the weak dependence is as followed. There still are few pores, impurity phases even in the relatively pure sample, which can reduce inter-grain current density. Besides, there are weak points, usually composition gradients of SmFeAsO phases, in most of superconducting circuits. When temperature or fields increase, these superconducting circuits with weak points are destroyed, and the inter-grain current density decreases. Though a few circuits, which are made up by clear boundaries and appropriate fluorine doped SmFeAsO phases, still exist even in high temperature or fields, the superconducting currents in these circuits, which can be called infiltration currents, are very small.

**Conclusions**

Polycrystalline $SmFeAsO_{1-x}F_x$ bulks with various compositions(x=0.15, 0.2, 0.3 and 0.4) were prepared by the conventional solid state reaction. The series of $SmFeAsO_{1-x}F_x$ were studied in terms of lattice constants, resistivity, magnetization, critical current density and microstructure. Our results show that critical transition temperature increases steadily with increasing fluorine content, and reaches its maximum of 53 K at x=0.4, which is different from the results of previous study. A study on Tc, lattice constants and $\rho_{60K}$ of various samples indicates that the shrinkage of a-axis seems to be the main reason for the rising of Tc and the change of c-axis has a large effect on carrier density. A global critical current density of $1.1\times10^4$ A/cm$^2$ at 5 K in self field was achieved in the purest sample. An inter-grain Jc of $3.6\times10^3$ A/cm$^2$ at 5 K in self field were estimated, which shows a weak field dependence. We believe that composition gradients are responsible for the relatively low inter-grain Jc in fields, which can be significantly increased by improving the homogeneity of composition.

**Acknowledgements**

The authors thank Profs. Yong Zhao, Zizhao Gan, Liye Xiao, Haihu Wen and Liangzhen Lin for their help and useful discussion. This work is partially supported by the Natural Science Foundation of China (Contract Nos. 50572104 and 50777062) and National ‘973’ Program (Grant No. 2006CB601004).

## Captions

Figure 1 (a) X-ray diffraction patterns of $SmFeAsO_{1-x}F_x$ samples. (b) shows the lattice constants with the error bar as a function of starting composition.

Figure 2 (a) Temperature dependence of resistivity for $SmFeAsO_{1-x}F_x$ samples. (b) The variations of $Tc_{onset}$ and $\rho_{60K}$ with x.

Figure 3 Temperature dependence of DC magnetic susceptibility of the sample with the highest Tc (x=0.4)

Figure 4 (a) SEM images of the cross sections of the $SmFeAsO_{0.8}F_{0.2}$ bulk, (b) High magnification images of the cross-section, (c) SEM images of the powder sample, (d) EDX analysis of a plate-like grain in the cross-section of the $SmFeAsO_{0.8}F_{0.2}$ bulk.

Figure 5 (a) Normalized magnetic hysteresis loops of $SmFeAsO_{1-x}F_x$ bulk at 5 K. (b) Critical current density calculated using Bean model.

Figure 6 (a) Normalized magnetization hysteretic loops of the $SmFeAsO_{0.8}F_{0.2}$ bulk and powders ground from the bulk at 5 K. (b) Inter-grain Jc calculated from the difference in magnetization between bulk and powders ground from the bulk.

Figure 7 EDX mapping images of the SmFeAs $O_{0.8}F_{0.2}$ bulk.

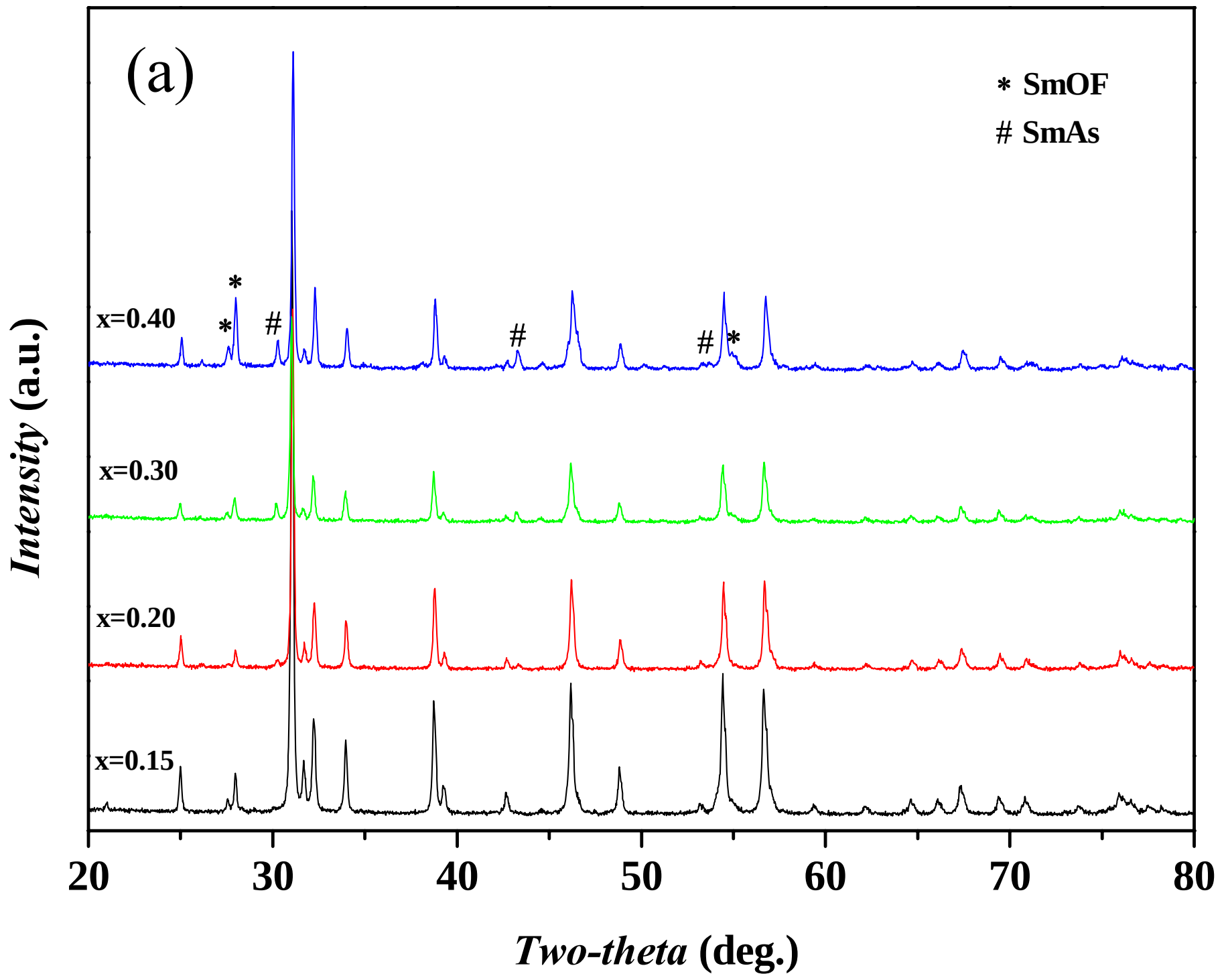

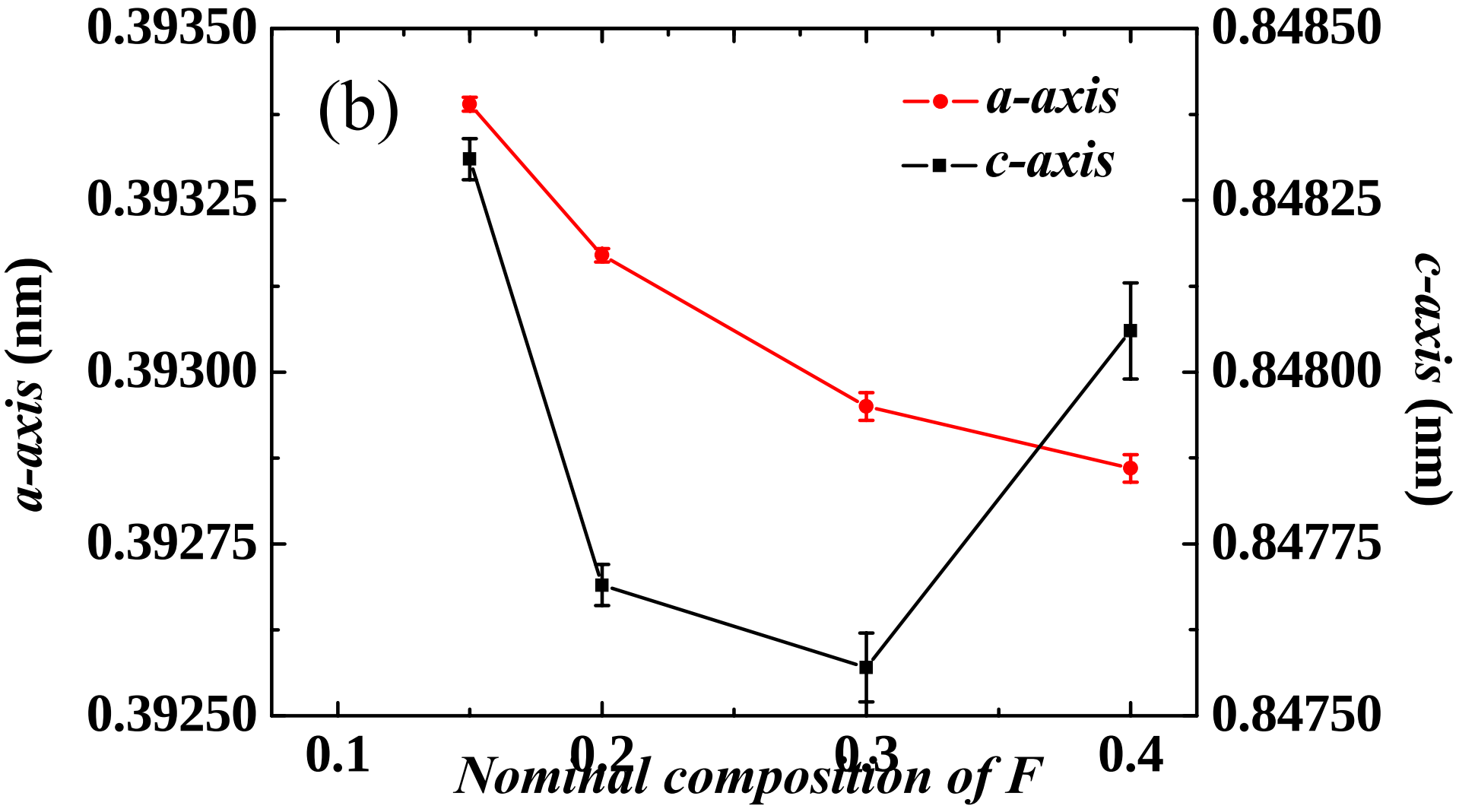

Figure 1 Wang et al.

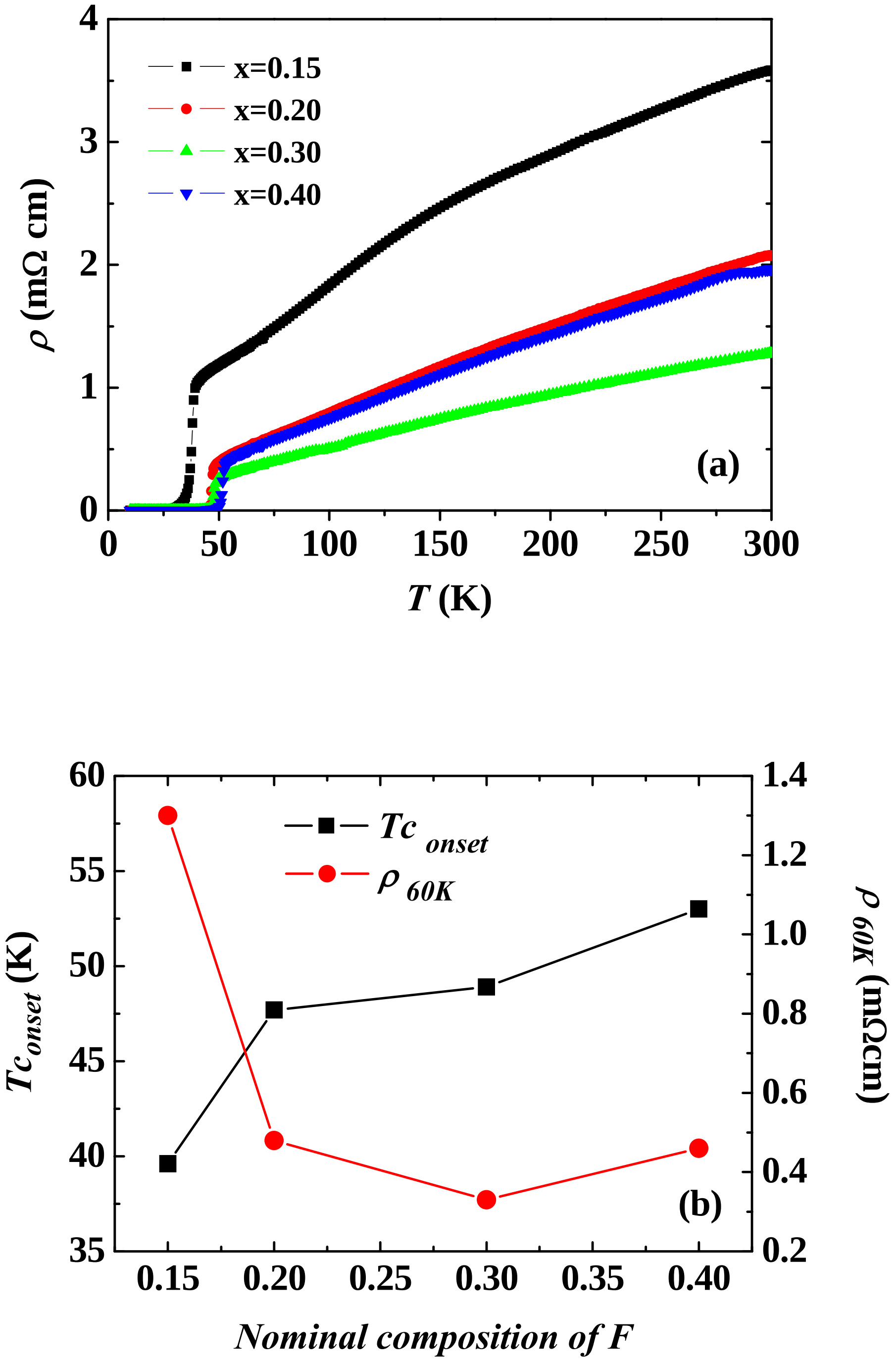

Figure 2 Wang et al.

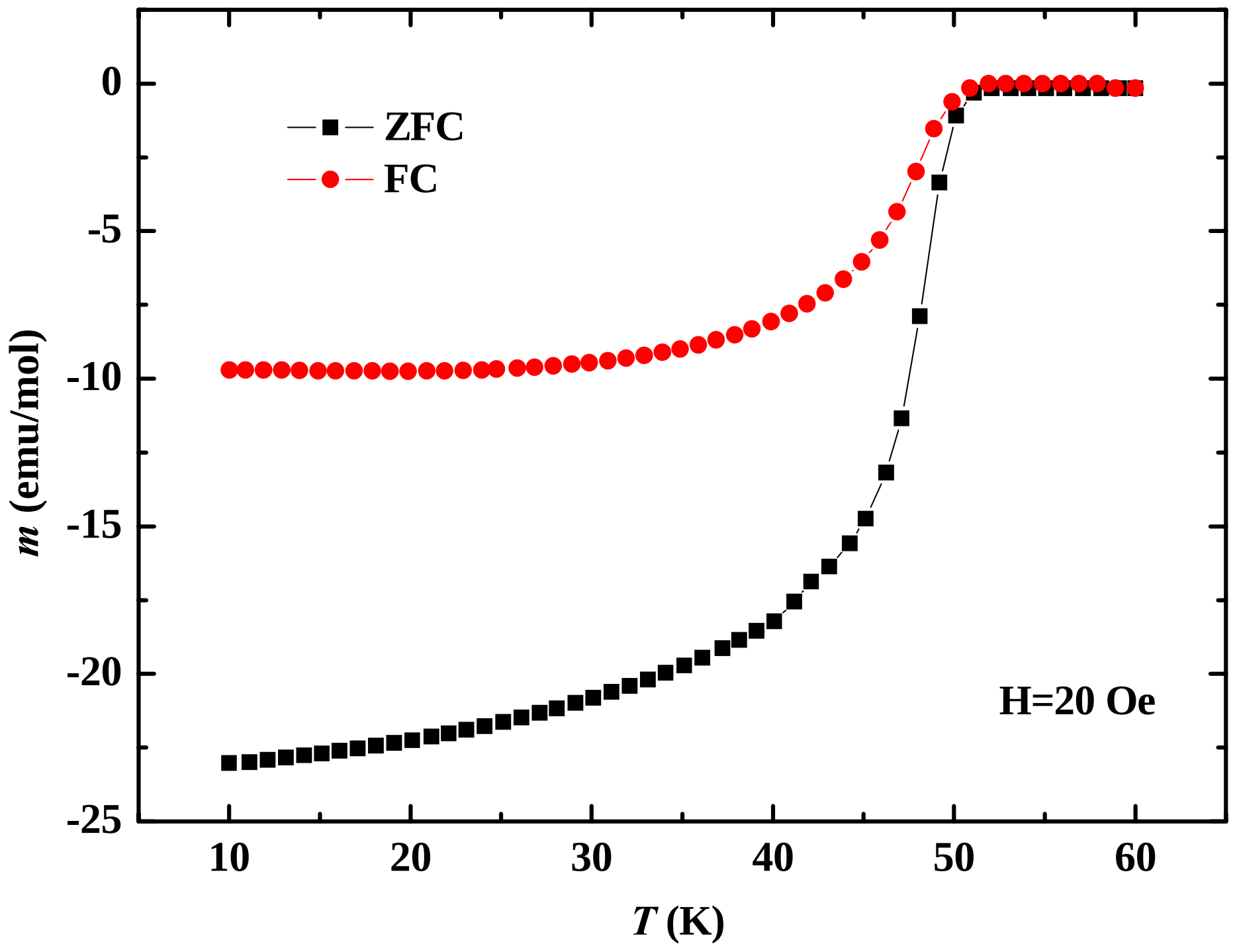

Figure 3 Wang et al.

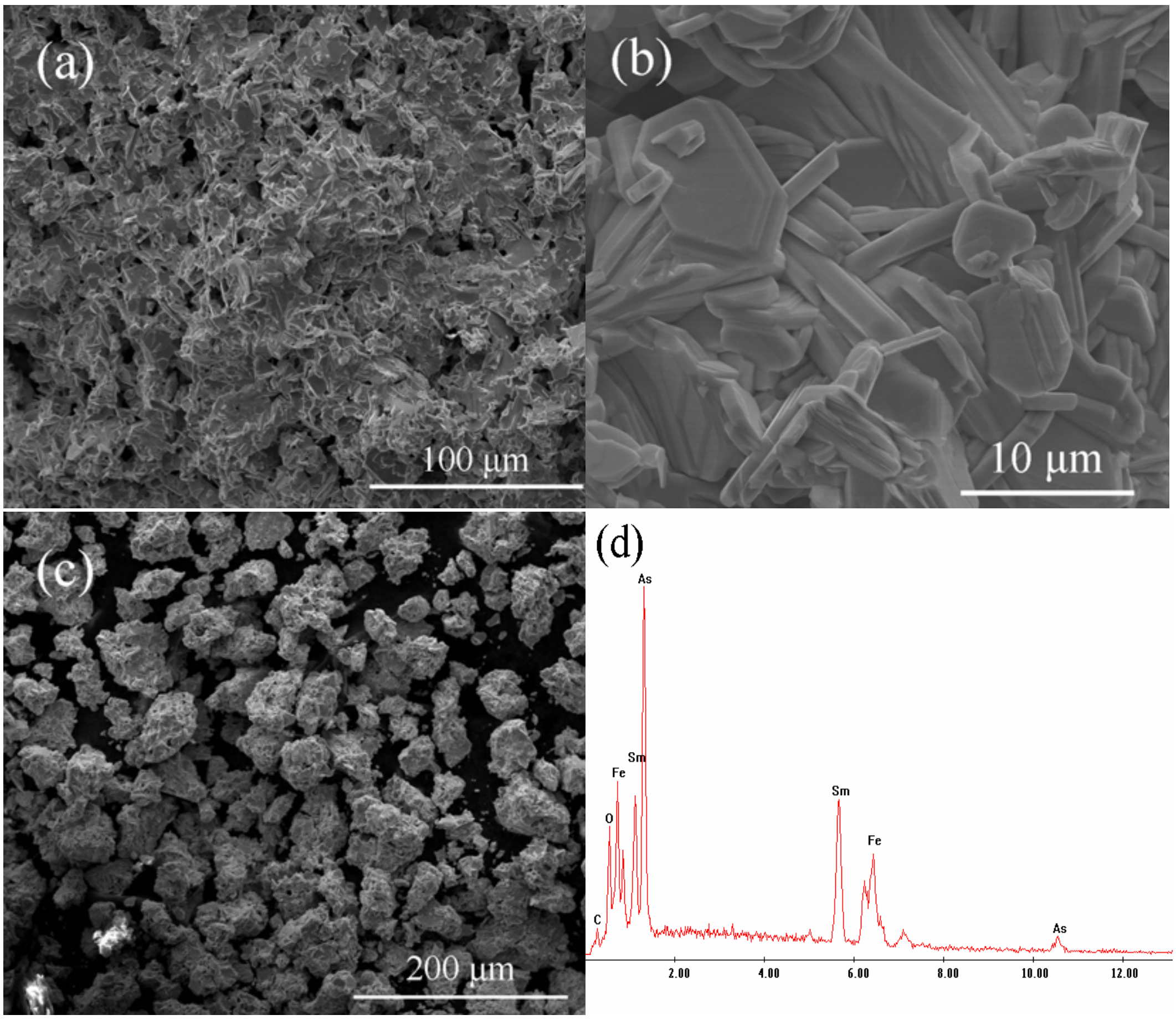

Figure 4 Wang et al.

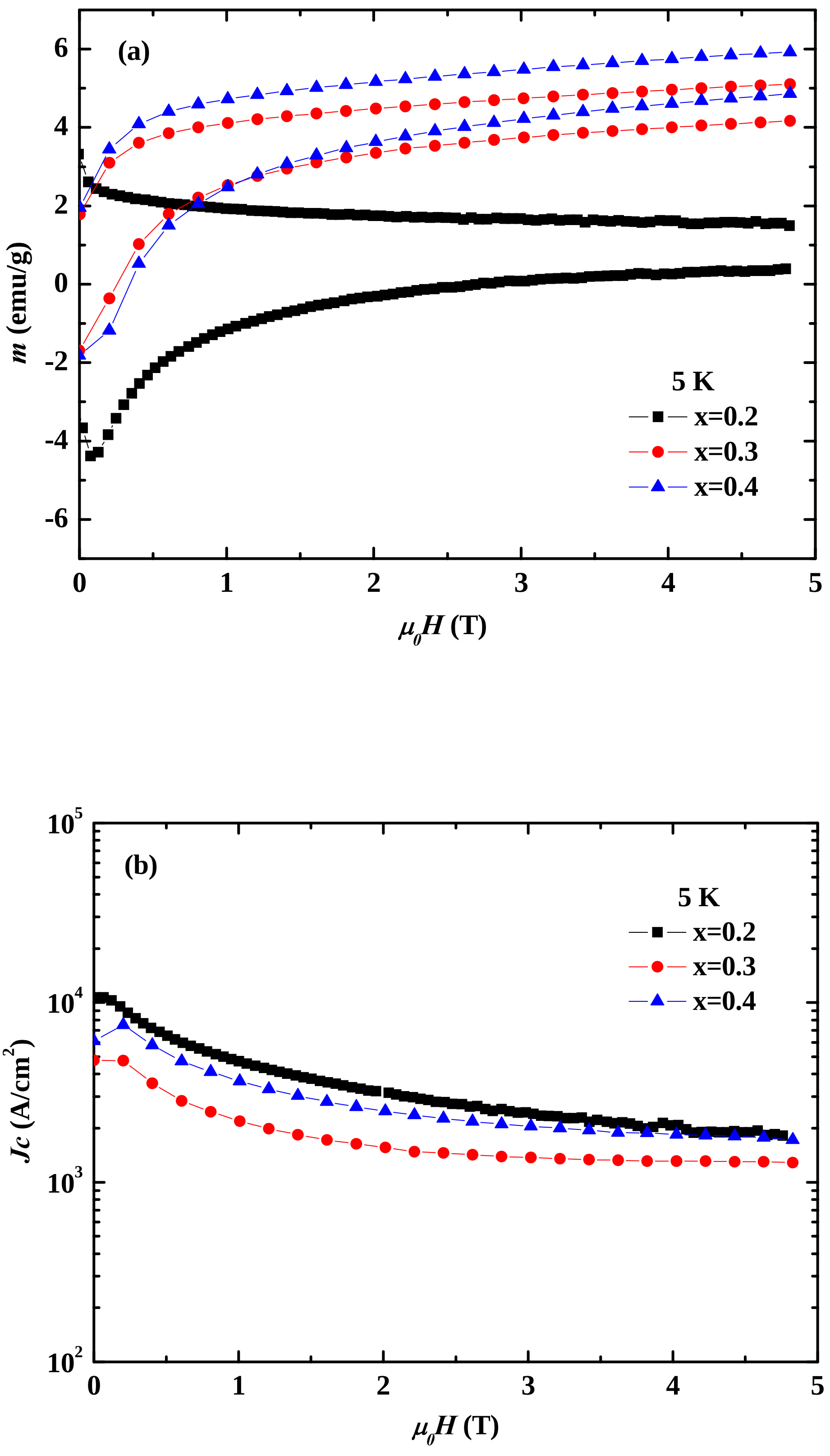

Figure 5 Wang et al.

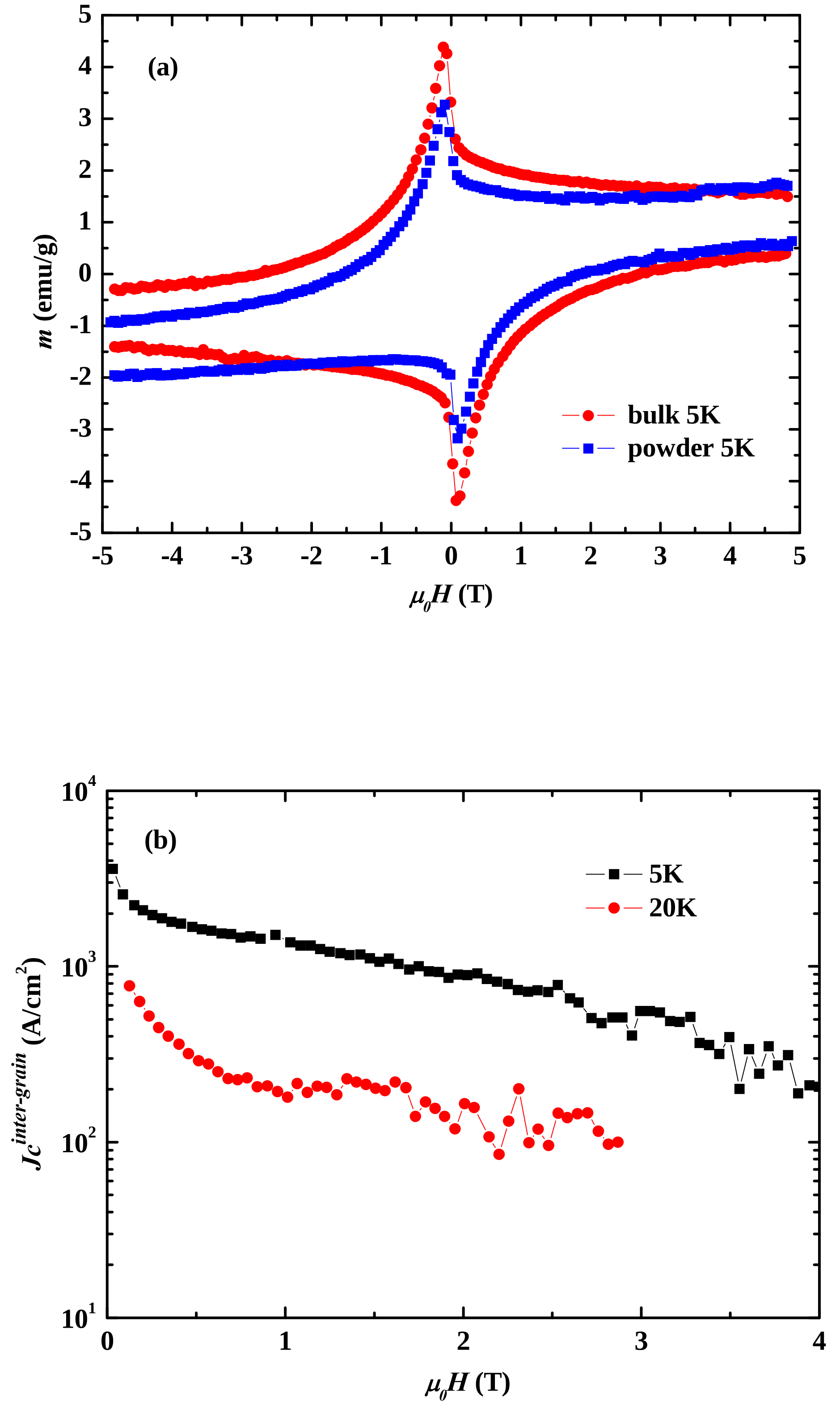

Figure 6 Wang et al.

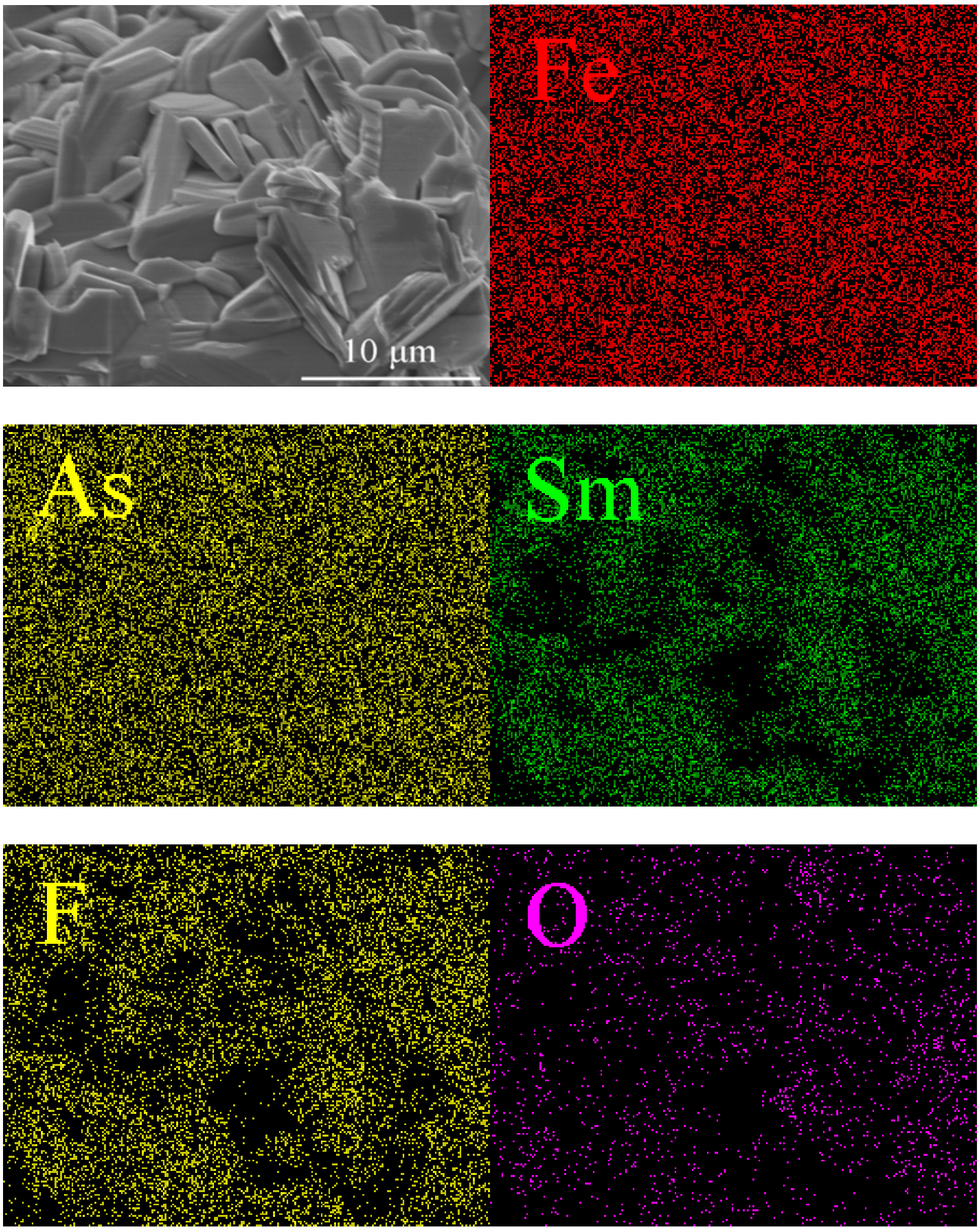

Figure 7 Wang et al.